# The Design of Circuit-Measuring Collaborative Learning System with Embedded Broker

Fu-Chien Kao, Siang-Ru Wang and Ting-Hao Huang

Computer Science and Information Engineering,
Da-Yeh University, Taiwan

**ABSTRACT**

Recently, the academic community has been giving much attention to Cooperative Learning System, a group learning method combined with pedagogy and social psychology. It allows group members to gain knowledge through collaborations and interactions. Nowadays, most Internet cooperative learning systems are designed to provide students mainly with a convenient online environment to study theoretical courses but rarely with an online environment to operate practical instruments. Hence, this paper designed a 3D online cooperative learning system for operating virtual instruments with circuit-measuring function. By integrating with Virtual Reality, Remote Control Parameter Transmission and embedded system techniques, this system gives learners not only a cooperative learning environment via networking to jointly operate the 3D virtual instruments (for example, multi-meters, power supplies and oscilloscopes) but also the functions of instant messages and 3D puzzles to interact with one another. Therefore, learners can effectively improve learning interests and results.

## 1. Introduction

In the traditional learning, the learning process of the students is similar to photocopy. They catch the knowledge transferred by teachers. The teachers teach in one-way. Although the questions and suggestions from the students are important, the influence is in one-way. The teachers have been acting the models all the time. Starting from the putting forward of the questions, analysis, reasoning to answer, all of them are dominated and demonstrated by the teachers and the students are passive. They observe and simulate the contents taught by the teachers and the learning results are not good. Under the traditional learning, the students are short of chance for discussion, communication and argumentation with colleagues. Therefore, they have no idea on how the others consider on solving the same problem, how they can clarify the own views through the interaction with others and how the others organize the knowledge and concept from the communication and argumentation. The traditional learning becomes an accumulation of knowledge. The teachers are responsible for inculcating the results of the evolvement to the students. This learning activity does not only drab but it is not meaningful.

In general, there are there learning modes, including Competitive Learning, Individualistic Learning and Cooperative Learning (Huang, 1996). In the Competitive Learning, the students compete for the achievement with each other. They pursue the targets that can be achieved by a few students. Thus, the students may make every effort to win, or they may give up totally and escape from the reality. In the Individualistic Learning, the students are arranged to make effort independently to achieve individual targets. The students pursue for individual benefits, and they attach importance to their own efforts and results. They do not mind the learning and achievements of other colleagues. In the Cooperative Learning, they must help each other. They discuss, ask and answer questions mutually, and feedback in groups in order to achieve the individual and group targets. This does not only benefit the individual students but also benefits to the group members (Chen, 1997; Huang, 1996). In the process of Cooperative Learning, all members are benefited from the groups, and the learning usually happens when student are active and collaborate in solving a problem in a social environment (Laurillard, 2002). In fact, recent pedagogical research shows that learning is not simply knowledge assimilated with the help of a more knowledgeable person or mediated by a computer system, but also jointly constructed through solving problems with peers by a process of building shared understanding (Scardamalia & Bereiter, 2003). It is a kind of life community and is a learning trend that is suitable for team cooperation and innovation in the existing high-tech society.

There is a series of innovative teaching methods applied in educational domain recently. The innovative methods develop the education to reach a new era. The results from many researches (Zhang, 2003; Lin, 2005; Neo, 2006) also show that the teaching contents should interact with students regardless of the multimedia network teaching or the recent virtual reality auxiliary teaching system in order to reach the best learning performance. The computer is not the machine for data analysis only. From the designed digital teaching materials, it does not only promote the interests of the learners but also improves the learning efficiency. Furthermore, the learners all over the world can use the teaching materials easily through the network. Many domestic and foreign researchers on network Cooperative Learning wish to establish a more effective learning environment through the Internet in order to improve the learning ability and performance of the students (Shi, 2002; Chen, 2004; Dong, 2004). The existing =be classified into three types according to the tools and technologies of design. The first type is the network Cooperative Learning system that provides documents mainly (Johnson, 1994 ; Maier, 1994). This type of Cooperative Learning system discusses the design of the shared document database and the system infrastructure of the network Cooperative Learning. The second type is the Cooperative Learning system that provides video conferencing function. As per the name, it is suggested that the Cooperative Learning system mainly provides the mechanism for the members of the group to discuss face-to-face on network. The third type of Cooperative Learning system provides virtual reality environment (Maier, 1994). This type of Cooperative Learning system mainly provides a learning environment of simulated virtual reality to the members of the group. However, most of the existing network Cooperative Learning system a convenient network environment for the students to conduct online Cooperative Learning of theoretical courses. It is rarely to discuss how the Cooperative Learning environment of operation training of online-simulated instruments and the circuit current measurement can be established.

This paper has proposed a Web-based Cooperative Learning system for remote operation 3D virtual electronic instruments with circuit-measuring function. The system does not only provide the cooperative operation practice of simulated instruments on network for the learners, it also improves the interests of the learners in practical courses and the learning performance through the system functions including real time text chatting and discussion, and actual circuit measurement. The chapters of this article also include embedded system and applications, virtual reality, and system infrastructure implementation, the Cooperative Learning of virtual instruments operation and circuit measurement, and conclusions.

## 2. Embedded System and Applications

The "Embedded system" integrates the application of information software and hardware. The application of embedded system can be found everywhere, including the life facilities such as mobile phone, electric toy and video-audio instrument, and transportation system and automation in factories. The trend of the functions of common embedded systems is simplification. The software and hardware only include the modules required for specific functions. The peripheral products from different manufacturers are combined to become an intact embedded system according to the provided Intelligence Property (IP). Comparing with the common computers, the building cost can be reduced significantly. Besides, due to the domain of the application, most of them include the characteristics of miniaturization and low-power consumption. Since the system only includes the specific functions, the design of the system will be optimized to ensure the stability of the system (Kao, 2007).

2.1 Development Approach of Embedded System

Since the embedded system is developed for specific functions, the developing tools include In-Circuit Emulator (ICE), Development Board, Integrated Development Environment (IDE) and compiler (Microtime Computer, 2004). The Development Board is the design sample provided by the hardware manufacturers. The debugging mechanism in the central processing unit of the In-Circuit together with the pre-set messages of the compiler can simulate the program running on the hardware. The product developers may refer to their circuit design and integrate to the corresponding developing environment. Therefore, most of the Development Boards have been built with many peripheral function modules such as network card, seven-segment display, parallel port, LED and dip switch, for the convenient of product development and speedup the product development cycle. The application integrating the development environment and complier refer to the development software on the personal computer of the developer. Therefore, the system developers may design the working platforms and the corresponding applications of the embedded system on the personal computer, and complete the compiling. Through the parallel port or serial port, the corresponding software can be embedded to the target board and the cross-platform development process is finished. Since the common low-cost multimeter cannot communicate with computer, this study has made use of low-cost 8-bit 89c51 chip to develop Embedded RS-2332 Module to capture the data shown on the panel of the multimeter and transfer it to the computer (as shown in Figure 1).

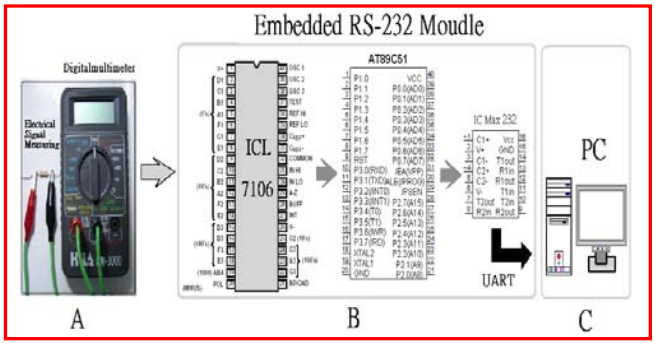

Figure 1 Embedded RS-232 module

## 2.2 Embedded RS-232 Module

Since most of the low-cost multimeters used by students do not include the serial communication function with the computer, this study makes use of 8-bit 89C51 single chip to implement an Embedded RS-232 Module. It is then integrated with the low-cost multimeter to replace the high-cost multimeter. The 89C51 single chip includes the following advantages: (1) the program is easy to learn (2) low-cost (3) simple circuit (4) small size. The infrastructure of the single chip is shown in Figure 2. The corresponding specification is as follows:

(1) CPU: the automatic controlled high performance 8-bit CPU is used to implement the whole operation of the computer.
(2) Program memory: the ROM or EPROM is used to store the program and the constant numbers. Different memories
    have different codes.
(3) Data memory: the RAM is used to store the data that is dynamic during the running of the program. The data can be
    accessed in the memory with the CPU. The data will disappear if the power is off.
(4) Oscillator: An oscillator is included in the MCS-51 series single chip. It will produce the clock for the whole system if
    it is connected with a quartz crystal.
(5) I/O pins: there are 32 input / output pins to use.
(6) Timer / Counter: the commands are used to set the 16-bit Timer or 16-byte Counter.

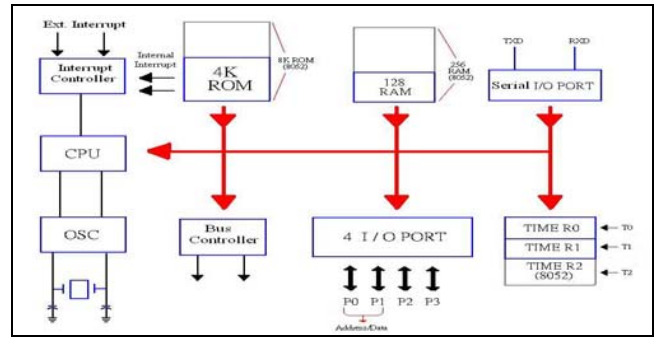

Figure 2 89C51 block diagram

## 2.3 Embedded Broker

As the network technology and applications are developed rapidly, large amount of servers for different network services are built. They include file server, web server and Email server. These servers are built on high-end servers. According to the functions built by the users, these powerful servers become the specific function servers. The hardware is very complicated and expensive, and it increases the building cost of the system directly. To build the network server by the embedded system, the relevant functions can be enhanced according to the requirement of the service. For example, the files operating performance and the space of the hard disks of the file server should be increased but the relatively low performance of CPU can be used. This can reduce the cost effectively and increase the competitiveness. This paper has adopted ARM7 development board (as shown in Figure 3) to develop embedded Broker to replace the expensive server.

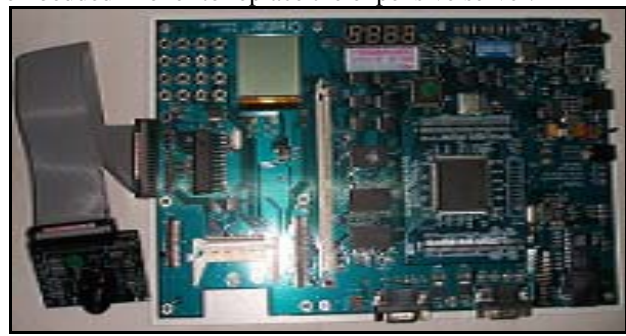

Figure 3 ARM7TDMI development board

## 3. Application of Virtual Reality Technology

The Virtual Reality (VR) is the term for science and technology domain. The virtual reality includes many components such as vision, hearing, and even sense of touch and sense of smell. The users can determine the path for browsing the information freely in the virtual reality.

Therefore, the virtual reality can be regarded as an intact multimedia system. The major characteristics of the virtual reality system are the interaction and the real time response. When it is applied on application, the users can operate the computers freely. They can observe the products in any angle and position. The virtual reality technology makes use of the computer drawing or image synthetic technology together with the virtual reality constructed by sound processing, sense of touch and sense of taste. In this virtual world, the 3D simulated object can be our famous things, or something that we cannot see, or a simulated imagined space. Virtual reality is an integrated technology in order to provide a higher level of man-machine interface (Kao, 2007). Dong (2004) has proposed three important topics for virtual reality. They are immersion, interaction and imagination (3I). These three topics are explained respectively as follows:

(1) Immersion:

The users can feel the scene of virtual reality directly produced by computer. The display system provides a simulated and First-person virtual world and scene. Besides, it can be controlled by the users directly. The users can really "immerse" in the world of virtual reality.

(2) Interaction

The users interact with the objects in the virtual reality world. For example, the users can walk in the virtual world, or wear the glove to catch the objects in the virtual world, or even interact with other users in the same virtual world.

(3) Imagination:

The virtual reality provides imagining space to the users. If the virtual reality technology is applied in the interacting environment of the virtual group, then multiply users can remote logon simultaneously, and share and interact in the same space, and meet the characteristics of virtual reality. This must bring better efficiency on interaction and communication approach to the virtual group. The users can talk and interact as if they are in the real world. The application and development of virtual reality technology on different professional domains are increasing gradually. One of the reasons is that the function of the software of virtual reality is enhanced. It is more important that the cost for virtual reality on personal computer (PC) is relatively low and it is widely accepted by all trades and professions. The 3D Webmaster and Virtools Dev 3.0 have been adopted in this study to develop the relevant virtual instruments.

The 3D Webmaster does not only provide the editing function for the 3D virtual object but it can also transform the DXF or VRML2.0 format to other 3D format. It can also export to VRML2.0 format and the audio file is mainly in wav format. The Direct 3D is supported and Z buffer is used for image processing so that the images look real and smooth. The system has no strict limitation to the hardware and it supports most of the peripheral devices of virtual reality. In addition, the SDK (Superscape Developers Kit) is applicable to compiling to the Dynamic Linking Libraries (.dll file) so that it can plug-in to the virtual scene and develop the own drivers. The virtual reality tool can develop online virtual reality and multiple users' connections on network. It can also process dynamic data link with other applications for information sharing. It provides an intact virtual scene to the users and they can browse the virtual scene from network after they have installed ActiveX on the browser. This software includes different editing tools, such as model editor, image editor, 3D scene and 3D virtual object editor. In order to increase the control of the 3D virtual object and flexibility of the interactive design of the 3D virtual object, the 3D Webmaster provides Superscape Control Language (SCL), which is similar to programming language C. Therefore, the users can write with the control language in the virtual scene designed by themselves (Superscape, 1996) as shown in Figure 4.

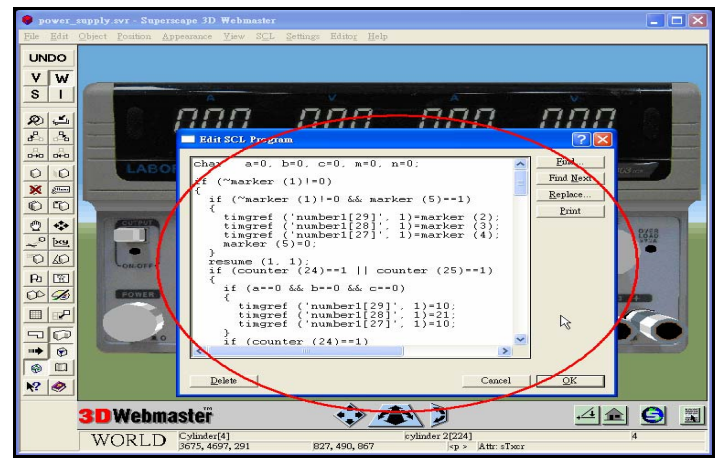

Figure 4 SCL program editor of 3D Webmaster

## 4. System Infrastructure and Implementation

The network technology is developed rapidly. For example, the web language starts from HTML, SMIL, VRML to XML. The media appeared starts from text, picture, and image to streaming media. The transmission channel starts from modem, ISDN, T1 lease line to ADSL. The change causes the revolutionary development on teaching method, teaching style and interactive mechanism. In order to recognize the characteristics of Cooperative Learning, the system of this study provides the functions including group real time discussion, real time interaction, operation of 3D virtual electronic instruments that include transmission technology of remote control parameters and circuit measurement.

### 4.1 System infrastructure

The functions of the system are divided into four major parts as shown in Figure 5.

(1) Cooperative Learning system: the learners can download the installation file from the link of the portal website. After the installation, the Cooperative Learning system can be entered.

(2) Cooperative Digital Learning: in the Cooperative Learning system, many items of Cooperative Digital Learning (system brief, 3D virtual lab, 3D puzzle that trains up initiative and interaction, 3D virtual multimeter and 3D virtual power supply) can be found. The learners can click and start Cooperative Digital Learning.

(3) Teaching material: after the learners have started the Cooperative Learning system, they can start learning from the digital teaching content. The users must study the operation of the electronic instruments first. Then they can make use of the real time discussion to practice operating the instruments and start Cooperative Learning with the group members.

(4) Circuit measurement: the learners of the same group can progress Cooperative Learning of multimeter with embedded RS-232 module for circuit measurement. The members of the same group can see the data of the circuit obtained by the measuring team from the 3D virtual instrument simultaneously.

### 4.2 Cooperative Learning of 3D virtual instruments with circuit- measuring function

The proposed cooperative practice environment with circuit measurement function sends the measured value from the multimeter Embedded RS-232 Module to the 3D virtual multimeters of the learners in the same group. Any learner can operate the own multimeter to progress cooperative practice. As shown in Figure 6, when the user sends the control command (or get the data from circuit measurement) to the 3D virtual instrument through the Operating Interface, it will be determined and processed by the embedded control program and a control parameter will be obtained. Then with the remote control parameter transmission technology, the embedded Broker transmits the control parameter to other learners of the same group so that the 3D virtual instruments of the learning partners can be updated simultaneously. The detailed transmission procedures of the remote control parameter are as follows:

(1) The user sends the control command (or receives the measured data of the circuit) on the Operating Interface. The obtained control parameter will be sent to 3D virtual instrument first and responded properly in accordance with the control parameter.

(2) The control parameter is sent to the embedded TCP/IP control function.

(3) The TCP/IP control function sends the control parameter to the Embedded Broker.

(4) Embedded Broker transmits the control parameter to the TCP/IP control functions of other users.

(5) Other users will send the received control parameter to the embedded control program.

(6) The embedded control program from other users then sends the control parameters to the 3D virtual instrument, and it responds properly in accordance to the control parameter.

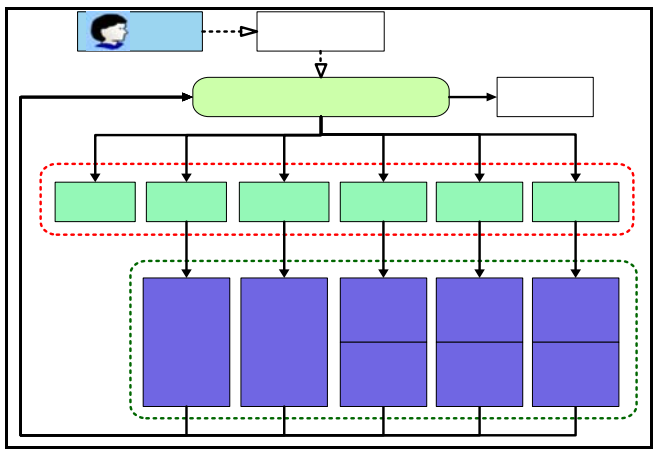

Figure 5 System infrastructure diagram

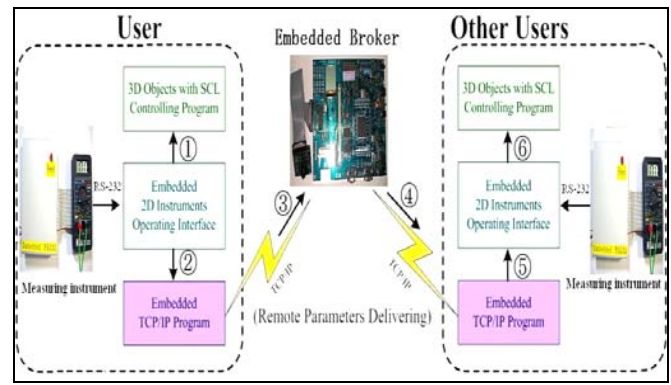

Figure 6 System operation procedures

In this way, the users can interact with the 3D virtual instruments and the display on the instruments can be synchronized among the users in the same group. The Embedded Broker is the backend of the whole system. When Embedded Broker is started, the main duty is to listen to the status of each user constantly. Once the status of a user is changed, Embedded Broker will receive the status changing information of the users. It will determine the groups of the learner immediately, and copies and sends the received information to the members of the same group in order to progress real time update of the status. Thus, the simulated operation of the instruments and the measured data of the circuit among the members in the same group will show the same result as shown in Figure 7 and Figure 8-10.

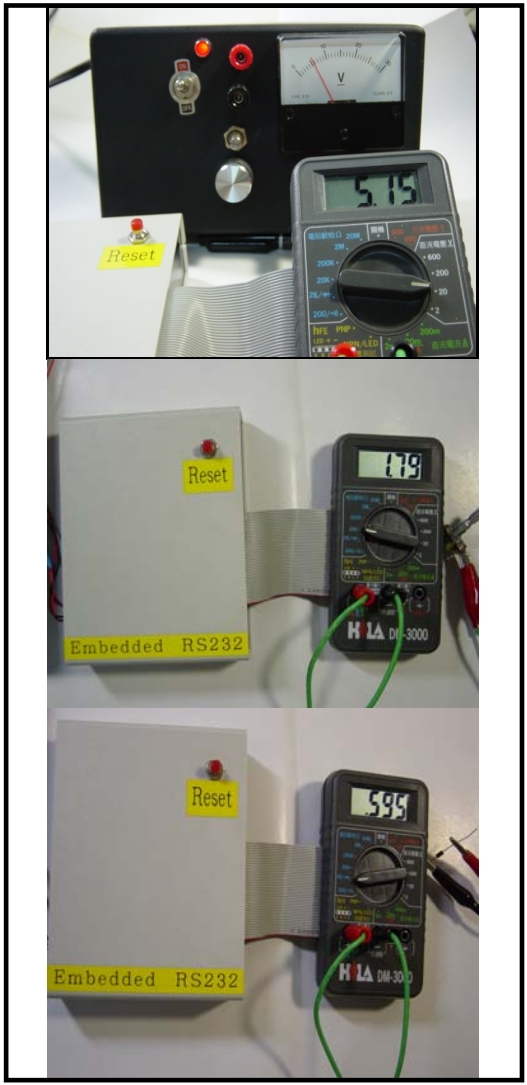

Figure 7 Circuit measurements by the multimeter

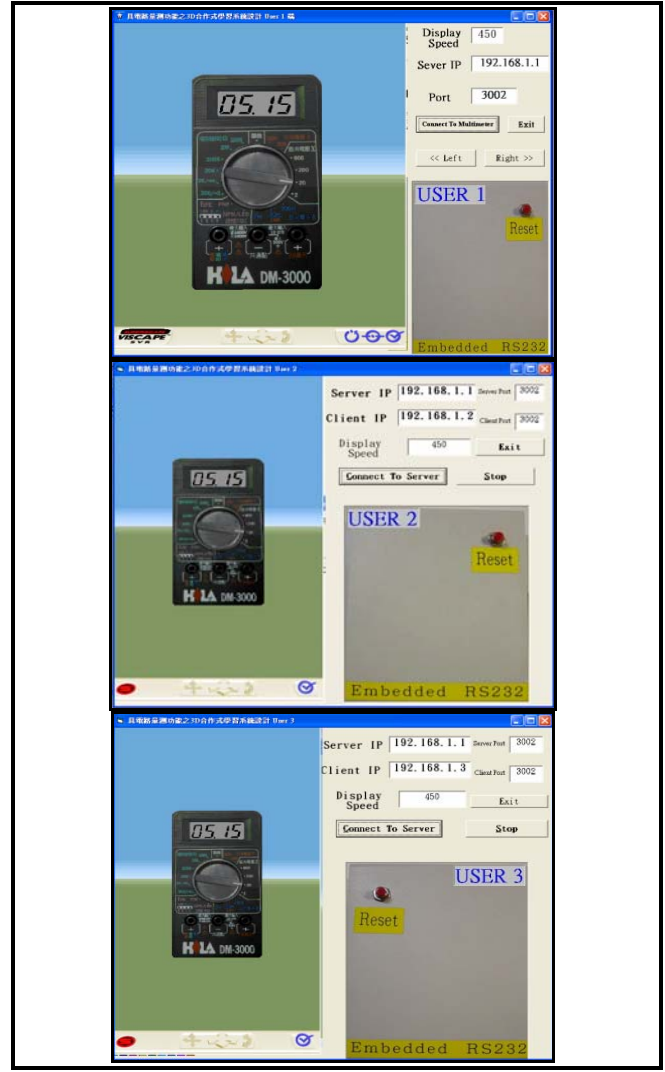

Figure 8 Voltage value shown from the 3D virtual instruments of the members in the same group

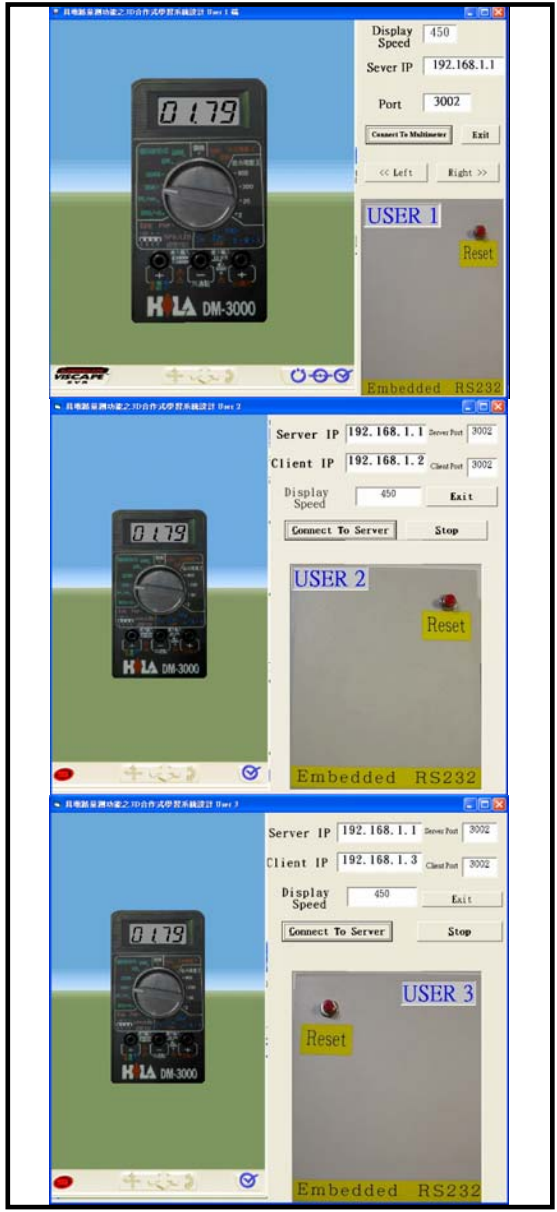

Figure 9 Resistance value shown from the 3D virtual instruments of the members in the same group

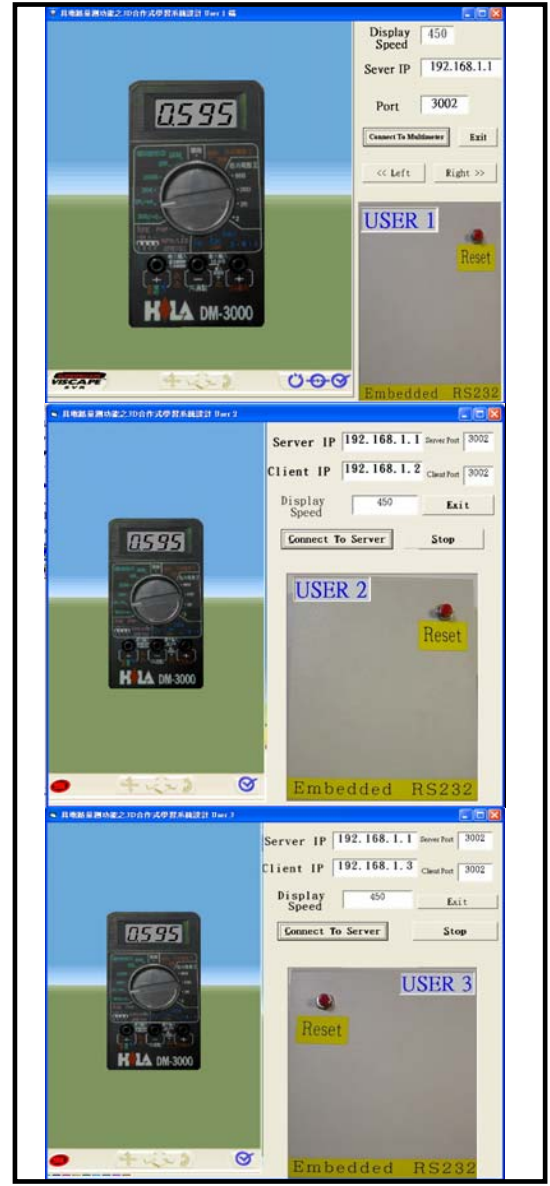

Figure 10 Diode measurement shown from the 3D virtual instruments of the members in the same group

### 4.3 Proposed 3D virtual instrument for Cooperative learning of simulated operation training

The designed 3D virtual instrument interface for Cooperative Learning of simulated operation training is shown in Figure 11. The functions of the system include login, group study, 3D virtual object operating interface and real time discussion. The corresponding function blocks are described below:

(1) The interactive 3D Virtual Instrument

(2) In the real time discussion, the users can transmit messages to the members of the group for real time discussion. The corresponding learning records can be saved for the teacher assessing the learning performance.

(3) User inputs the IP of the Embedded Broker and the assigned Port in order to connect to the Embedded Broker.

(4) After grouping, team study via networking available in addition to login learning.

(5) The users can make use of the virtual machine operating interface to operate the 3D virtual instrument.

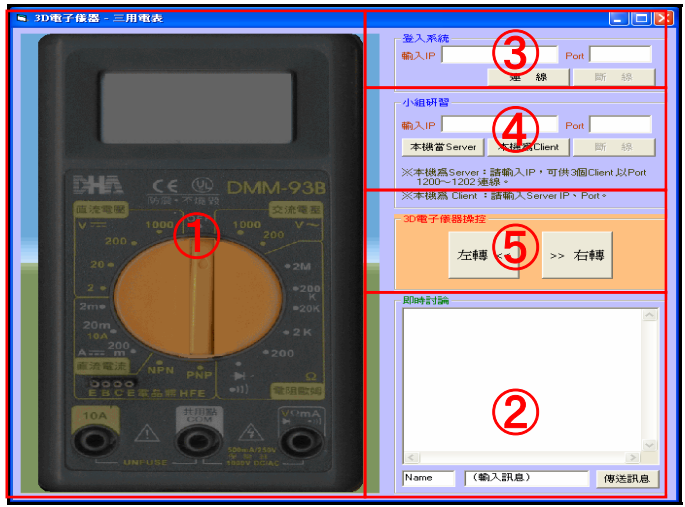

Figure 11 3D instruments in cooperative learning system

## 5. Performance Evaluation and Statistic Analysis

In the experiment, we adopts as the setting 40 learners with 3 users per group with 4 users in the 13th group, and connect into Embedded Broker to do cooperative learning by means of the cooperative learning materials offered by the system.

5.1 System Performance Testing

The hardware specification used by the 40 learners is Intel Pentium 3 Processor as CPU, main memory 512 MB, and Microsoft Window 2000 Professional Operating System. Learners can be allowed to make use of the Internet browser and go into the portal to download and install the system. The testing is focused on the performance of getting curriculum materials against increasing the number of user connections (one group per unit) to further analyze the transmission efficiency reached by the system. In order to retrieve the packets communicated between Embedded Broker and each learner's PC for analysis of data transmission, a professional network packet analyzer is utilized: Ethereal (Network Protocol Analyzer, version 0.10.10).

In Figure 12 and 13, the x-axis stands for the number of learners while the average response time for 3D virtual cooperative learning system (unit: ms, millisecond) is indicated by the y-axis. While 40 learners are using the system simultaneously, the response time for each learner will be different because the received status data calculated by averaging the sum of each learner's response time are orderly updated via Embedded Broker into each learner's PC. Under PC and Embedded Broker Architectures, the experimental results for the system's response time against increasing the number of learners are listed below:

(1) Under the system with PC Broker, while learning is done by grouping, the average response times for Group 1, 2, 3, and 4 are respectively "132ms," "144ms," "148ms," and "152~155ms," as illustrated in Figure 12.

(2) Under the system with Embedded Broker, while learning is done by grouping, the average response times for Group 1, 2, 3, and 4 are respectively "130ms," "141ms," "147ms," and "152~155ms," as illustrated in Figure 13.

From the above results, the proposed embedded Broker would also provide service capabilities similar to that of a high-end server system, and significantly reduce the system costs.

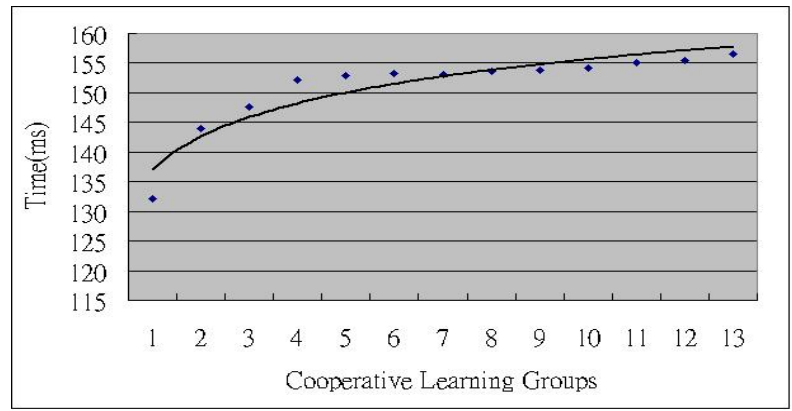

Figure 12 Average response time via PC Broker's group learning

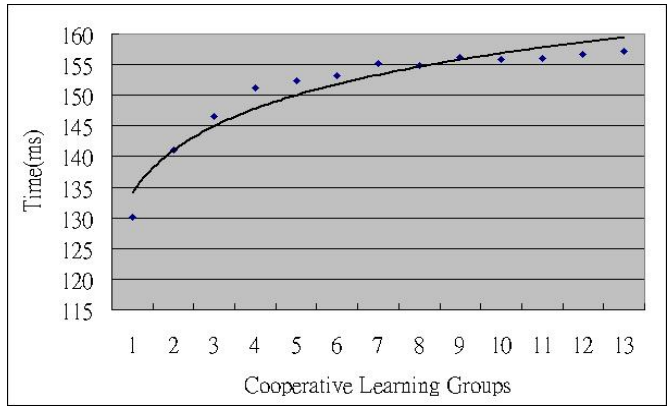

Figure 13 Average response time via embedded Broker's group learning

### 5.2 Questionnaire design and statistic analysis

The design of the questionnaire is using Likert-type scale. Each question have five options from not very agree, not agree, normal, agree, to strongly agree. The best result is 5 points, and the lowest is 1 point. The learners can choose the situation that they feel from question that they describe. The purpose of this survey is to analyze the actual use with co-operative practice system, and then make an assessment of learners progress the curriculum. We use online Assessment methods, and let 40 learners fill in the survey online after they use the cooperative learning system that is proposed by this research. From collected questionnaire data, the experts can analyze each question to related statistical results such as Table 1. From Table 1, the average grade is between 3.26 to 4.36. The statistic results show that the design of cooperative circuit measuring really improve the learning efficiency, and almost all students think that the hardware design of instrument with embedded RS-232 module is easy to operate.

Table 1 Statistic analysis

| Questionnaires | Average Score | Total Average |
|---|---|---|
| 1. The operations of virtual instruments really Helps the understanding of real instrument. | 3.5 | 3.80 |
| 2. The operation of the 3D cooperative type Instruments can really promote the positive Interdependence attitude each other among the Numbers of group. | 3.26 | |
| 3. The operation of the 3D cooperative type Instruments can promote the learning interest And learning efficiency. | 3.52 | |
| 4. The appearance design of the 3D cooperative Type instruments is very suitable for learning. | 3.63 | |
| 5. It is very simple that the cooperative operation of instruments with embedded RS-232 module. | 4.22 | |
| 6. The operating interface is suitable and easy use for learners. | 4.08 | |
| 7. The cooperative circuit measuring of real VOM instrument can really improve the learning efficiency. | 4.36 | |

## 6. Conclusions

Most of the existing network Cooperative Learning system provides a convenient network environment for the students to conduct online Cooperative Learning of theoretical courses. It is rarely to discuss how the Cooperative Learning environment of operation training of online-simulated instruments and the circuit measurement can be established. This causes the existing network Cooperative Learning system cannot implement the characteristics of Cooperative Learning in the learning environment on the Internet.

This study has broken through the traditional "lab operation" practice mode. The Cooperative Learning environment of the instrument practice is implemented in the Internet. The proposed Cooperative Learning system of 3D virtual instruments with circuit-measuring function integrates with the Virtual reality technology, embedded system technology and the transmission technology of the remote control parameter. The Cooperative Learning environment for operation of instruments on Internet is established completely. It provides the Cooperative Learning practice of circuit measurement for the learners to operate online anytime. Besides, the simulating training provided by the 3D virtual instruments, and the online grouped discussion do not only increase the interests of the learners in practice courses and train up their initiative and interaction on Cooperative Learning. This also decreases the damage rate of the traditional instruments and reduces the purchasing cost of school facilities. Besides, the proposed embedded Broker would also provide service capabilities similar to that of a high-end server system, and significantly reduce the system costs.

**Acknowledgment**

This paper has been supported in part by the grant NSC95-2520-S-212-001-MY3 from the National Science Council of Taiwan.